\documentclass[11pt]{article}
\def\la{\langle}
\def\ra{\rangle}
\def\beq{\begin{equation}}
\def\eeq{\end{equation}}
\def\be{\begin{eqnarray}}
\def\ee{\end{eqnarray}}

\def\k2av{\la k_T^2\ra}
\newcommand{\f}[2]{\frac{#1}{#2}}
\newcommand{\dd}{ {\textrm d}}

\begin{document}
%\title{Testing Nuclear Reaction Mechanisms \\ by Cronin Effect }
\title{Nuclear Reaction Mechanisms and the Cronin Effect }
\author{
G.G. Barnaf\"oldi$^{1,2}$, G. Papp$^{3}$, P. L\'evai$^{1}$
and G. Fai$^{4}$  \\[1ex]
$^{1}$ RMKI Research Institute for Particle and Nuclear Physics, \\
\ \ \  PO Box 49, Budapest, 1525, Hungary\\
$^{2}$ Laboratory for Information Technology, E{\"o}tv{\"o}s University \\ 
\ \ \  P{\'a}zm{\'a}ny P. 1/A, Budapest 1117, Hungary \\ 
$^{3}$ Department for Theoretical Physics,      
       E{\"o}tv{\"o}s University \\ 
\ \ \  P{\'a}zm{\'a}ny P. 1/A, Budapest 1117, Hungary \\ 
$^{4}$ Center for Nuclear Research, Department of Physics,  \\
\ \ \  Kent State University, Kent, OH 44242
}
\date{Presented at 10\textsuperscript{th} International Conference on 
Nuclear Reaction Mechanism \\ 10\textsuperscript{th} June 2003, Varenna, Italy }

\maketitle

\begin{abstract}
The influence of nuclear multiscattering and shadowing on pion spectra is 
investigated in $pA$ collisions from CERN SPS to RHIC energies. 
The calculations are performed in a next-to-leading order (NLO)
pQCD-improved parton model, including intrinsic partonic transverse momentum
distributions. The nuclear modification of the pion spectra (Cronin effect)
is considered at different targets in a wide energy range. Theoretical 
predictions are displayed for planned $pA$ experiments at CERN SPS 
and recent $dAu$ experimental results are analysed at RHIC.

%\vspace{.2cm}
%\keyword{pQCD, intrinsic $k_T$, pion production, $K$ factor, Cronin effect}
%\PACS{see {\tt http://www.aip.org/pubservs/pacs.html}}
%\PACS{24.85.+p, 13.85.Ni, 13.85.Qk, 25.75.-q; 25.75.Dw; 25.75.Nq; 12.38.Bx }
%\noindent {\em PACS numbers:} 
\end{abstract}

\section{Introduction}

%Closely after the Cronin effect was discovered in CERN 
%\cite{Cron75,Antr79,e605}, several theoretical investigations were launched to
%reproduce this $\sim 20 -40\%$ difference in particle 
%production between  proton-nucleus ($pA$) or 
%nucleus-nucleus ($AA$) collisions and scaled proton-proton ($pp$)
%one, where the scaling factor is the {\it number of binary collisions}.  

The Cronin effect was discovered in the energy 
range $20$ GeV $\leq \sqrt s \leq 40$ GeV at 
FERMILAB~\cite{Cron75,Antr79,e605}. The effect is a $\sim 20 -40\%$ 
enhancement in particle production in proton-nucleus ($pA$)
collisions compared to scaled proton-proton ($pp$) ones, where the 
scaling factor is the {\it number of binary collisions}. 
Recent RHIC experiments at $\sqrt s = 200$ GeV raised the interest on    
this nuclear effect in nucleus-nucleus 
($AA$) collisions. Theoretical investigations were launched to 
understand and reproduce the Cronin peak~\cite{Wang01,Wong98,YZ02}.     

Theoretical models were based on perturbative Quantum Cromodynamics (pQCD).   
General problem with pQCD models, that they underestimate the experimental 
data by $\sim 50 -100 \% $ at usual scales. Changing scales could  
solve this problem, however larger scale causes an increase in the lower 
limit of applicability of pQCD calculations over the region of interest 
$3 $ GeV$\leq p_T \leq 6$ GeV. Introducing the so called intrinsic transverse
momenta ($k_T$), one can reproduce the experimentally measured pion ($\pi$) 
spectra in $pp$ collisions within $ \sim 10-20 \%$. The relevant parameter of these 
calculations is $\langle k_T^2 \rangle $, the width of the intrinsic partonic 
transverse momentum distribution which depends on the c.m. 
energy ($\sqrt s$), at fix scales. 

Here pion production is investigated  in a wide (high-)energy 
range from CERN SPS up to RHIC ($20$ GeV $\leq \sqrt s \leq 200$ GeV). 
Although this analysis can be revisited in case of other 
hadrons\cite{YZ02,XZ03}, however due to the poor statistic of 
experimental data, in this paper we would like to reproduce only pionic 
data. Our calculation is performed in next-to-leading order (NLO), 
and compared to earlier leading order (LO) results~\cite{YZ02}.    

\section{Theoretical Background for $pp$ collisions}

First of all let us start with the invariant cross section for pion production 
in a $pp$ collision, which can be described in the $k_T$-enhanced NLO
 pQCD-improved parton model on the basis of the factorization theorem as a 
convolution\cite{pgNLO}:
\begin{eqnarray}
\label{hadX}
 E_{\pi}\f{\dd \sigma^{pp}}{\dd ^3p_\pi} &=&
        \f{1}{S} \sum_{abc}
  \int^{1-(1-V)/z_c}_{VW/z_c} \f{\dd v}{v(1-v)} \ 
  \int^{1}_{VW/vz_c} \f{ \dd w}{w} 
  \int^1 {\dd z_c} \nonumber \\
  && \ \int {\dd^2 {\bf k}_{Ta}} \ \int {\dd^2 {\bf k}_{Tb}}
        \, \, f_{a/p}(x_a,{\bf k}_{Ta},Q^2)
        \, f_{b/p}(x_b,{\bf k}_{Tb},Q^2) \cdot
   \nonumber \\
&& \cdot  
 \left[
 \f{\dd {\widetilde \sigma}}{\dd v} \delta (1-w)\, + \,
 \f{\alpha_s(Q_R)}{ \pi}  K_{ab,c}(s,v,w,Q,Q_R,Q_F) \right] 
 \f{D_{c}^{\pi} (z_c, Q_F^2)}{\pi z_c^2}  \,\,   ,
\end{eqnarray}
where we introduced factorised 3-dimensional parton distribution functions (PDFs), 
\be
f(x,{\bf k}_{T},Q^2) \,\,\,\, = \,\,\,\, f(x,Q^2) \cdot g({\bf k}_{T}) \ .
\ee
Here, the function $f(x,Q^2)$ represents the standard 1-dimensional NLO PDF  
as a function of momentum fraction of the incoming parton $x$ at factorization 
scale $Q$, $\dd {\widetilde \sigma}/ \dd v$ represents the Born cross section 
of the partonic subprocess $ab \to cd$, $K_{ab,c}(s,v,w,Q,Q_R,Q_F)$ is the 
corresponding higher order correction term, and the fragmentation function 
(FF), $D_{c}^{\pi}(z_c, Q_F^2)$, gives the probability for parton $c$ to 
fragment into a pion with momentum fraction $z_c$ at fragmentation scale $Q_F$.
We use the conventional proton level ($S,V,W$) and parton level ($s,v,w$)
kinematical variables of NLO calculations (see 
Ref.s~\cite{pgNLO,Aversa89,Aur00}).

In this analysis we consider fixed scales: the factorization and the 
renormalization scales are connected to the momentum of the intermediate 
jet, $Q=Q_R=\kappa\cdot p_q$ (where $p_q=p_T/z_c$), while the fragmentation 
scale is connected to the final hadron momentum, $Q_F=\kappa \cdot p_T$.
The value of $\kappa$ is $ \sim {\cal O} (1)$.  

For our NLO calculations we are using a  
3-dimensional PDF which is a product of the standard
PDF (at its $Q$ scale) and a 2-dimensional initial transverse-momentum    
distribution, $g({\bf k}_T)$ of partons containing its "intrinsic $k_T$" 
parameter as in Refs. \cite{YZ02,pgNLO,Wang01,Wong98}. We demonstrated the
success of such a treatment at LO level in Ref. \cite{YZ02}, and a 
$K_{jet}$-based NLO calculations in \cite{Bp02,bgg}.
In our phenomenological approach the transverse-momentum distribution
is described by a Gaussian,
\beq
\label{kTgauss}
g({\bf k}_T) \ = \f{1}{\pi \la k^2_T \ra}
        e^{-{k^2_T}/{\la k^2_T \ra}}    \,\,\, .
\eeq
Here, $\langle k_T^2 \rangle$ is the 2-dimensional width of the $k_T$
distribution and it is related to the magnitude of the
average transverse momentum of a parton
as $\langle k_T^2 \rangle = 4 \langle k_T \rangle^2 /\pi$.

In present article we use the LO GRV\cite{GRVLO}, and the NLO 
MRST(cg)\cite{MRST01} PDFs. For FF we apply most recent KKP 
parameterizations~\cite{KKP}, both in LO and NLO cases.
These PDF and FF sets can be applied down to very small scales 
($Q^2\approx 1.25$ GeV$^2$), thus we have got results at relatively 
small transverse momenta, $p_T \geq 2$ GeV at our fix scales. The details of the 
NLO calculations can be found in Ref.~\cite{pgNLO}

In our earlier work\cite{YZ02} we determined the $p_T$-independent 
intristic-$k_T$ values for the $3$ GeV $ \leq p_T \leq 6$ GeV  momentum
window and investigated their dependence on c.m. energy at fixed 
scales\cite{Bp02}. We found that the reproduction of the Cronin effect requires 
$\langle k_T^2 \rangle \approx 2$ GeV\textsuperscript{2}.  
The di-jet production data at ISR energies yield a similar value for 
$\k2av$~\cite{Angelis80}. 
Recent measurements of jet-jet correlations in $pp$ collision at RHIC 
energies\cite{STARdijet,PHENIXdijet} can clarify the properties of 
the transverse component of the PDFs. 

Our description of pion production in $pp$ collision can be extended
for $pA $ and $AA$ collisions.

%%%%%%%%%%%%%%%%%%%%%%%%%%%%%%%%%%%%%%%%%%%%%%%%%%%%%%%%%%%%%%%%%

\section{Nuclear Effects in $pA$ and $AA$ Collisions}

Proton-nucleus and nucleus-nucleus collisions can be described by 
including collision geometry, saturation in 
nucleon-nucleon ($NN$) collision number, and shadowing inside the nucleus. In 
the frame of Glauber picture, the cross section of pion production in 
nucleus-nucleus collision can be written as an integral over impact 
parameter $b$:
\beq
\label{dAuX}
  E_{\pi}\f{\dd \sigma_{\pi}^{AA'}}{ \dd ^3p} =
  \int \dd ^2b \, \dd ^2r \,\, t_A(r) \,\, t_{A'}(|{\bf b} - {\bf r}|) \cdot
  E_{\pi} \,    \f{\dd \sigma_{\pi}^{pp}(\k2av_{pA},\k2av_{pA'})}
{\dd ^3p}
\,\,\, ,
\eeq
where $pp$ cross section on the right hand side represents
the cross section from eq. (\ref{hadX}), but with an increased widths  
compared to the original transverse-momentum distributions (\ref{kTgauss}) in
$pp$ collisions, as a 
consequence of nuclear multiscattering (see eq. (\ref{ktbroadpA})).  
Here $t_{A}(b) = \int \dd z \, \rho_{A}(b,z)$ is the nuclear thickness 
function (in terms of the density distribution of nucleus $A$, $\rho_{A}$), 
normalized as $\int \dd ^2b \, t_{A}(b) = A$. 
For small size nucleus we used sharp sphere approximation, while for 
larger nuclei Wood-Saxon formula were applied.

The initial state broadening of the incoming parton's distribution function is 
accounted for by an increase in the width of the Gaussian parton transverse 
momentum distribution in eq. (\ref{kTgauss}):
\beq
\label{ktbroadpA}
\k2av_{pA} = \k2av_{pp} + C \cdot h_{pA}(b) \ .
\eeq
Here, $\k2av_{pp}$ is the width of the transverse momentum distribution
of partons in $pp$ collisions, $h_{pA}(b)$ describes the number of 
{\it effective} $NN$ collisions at impact parameter $b$, which impart an 
average transverse momentum squared $C$. The effectivity function 
$h_{pA}(b)$ can be written in terms of the number of collisions suffered 
by the incoming proton in the target nucleus, 
$\nu_A(b) = \sigma_{NN} t_{A}(b)$, where $\sigma_{NN}$ is the inelastic 
$NN$ cross section:
\begin{equation}
\label{hpab}
  h_{pA}(b) = \left\{ \begin{array}{cc}
                \nu_A(b)-1 & \nu_A(b) < \nu_{m} \\
                \nu_{m}-1 & \mbox{otherwise} \\
        \end{array} \right.\ .
\end{equation}
The value $\nu_{m}= \infty$ corresponds to the case where all possible
semihard collisions contribute to the broadening. We have  
found that for realistic nuclei the maximum number of 
semihard collisions is $3 \leq \nu_{m} \leq 4$.

The determination of the factor $C$ and of $\nu_A(b)$ is in
progress in a systematic analysis in NLO~\cite{NLOsys}. Our preliminary 
results confirm the findings of Ref.~\cite{YZ02}, where the systematic 
analysis of $pA$ reactions was performed in LO and the characteristics
of the Cronin effect were determined  at LO level. Following 
Ref.~\cite{YZ02}, we assume that only a limited number of semi-hard 
collisions (with $\nu_m = 4$) contributes to the broadening,
average momentum squared imparted per collision is $C =$ 0.4 GeV$^2$. 

Furthermore, the PDFs are modified in the nuclear environment by the 
``shadowing'' effect\cite{Shadxnw_uj,EKS,HKM}. This effect and isospin 
asymmetry are taken into account on average using a scale independent 
parameterization of the shadowing function $S_{a/A}(x)$ adopted from 
Ref.~\cite{Wang01}:
\begin{equation}
f_{a/A}(x,Q^2) =
S_{a/A}(x) \left[\frac{Z}{A} f_{a/p}(x,Q^2) + \left(1-\frac{Z}{A}\right)
  f_{a/n}(x,Q^2) \right]   \,\,\,\,  ,
\label{shadow}
\end{equation}
where $f_{a/n}(x,Q^2)$ is the PDF for the neutron and $Z$
is the number of protons.
In the present work, we display results obtained with the 
EKS parameterization~\cite{EKS}, and with 
the updated HIJING parameterization\cite{Shadxnw_uj}. The former has an 
antishadowing feature, while the later incorporates different 
quark and gluon shadowing, and 
has an impact-parameter dependent and an impact-parameter independent version.
The impact-parameter dependence is taken into account
by a term $\propto (1-b^2/R_A^2)$, which re-weighs the shadowing
effect inside the nucleus.

We note, that for the deuteron ($d$), one could use a simple 
superposition of a $pAu$ and a $nAu$ collision, or a distribution for 
the nucleons inside the deuteron.  For a first orientation we follow 
Ref. \cite{Vitevlo,dAu} in this regard, and apply a hard-sphere 
approximation for the deuteron with $A=2$ %for estimating the nuclear effects, 
taking no nuclear effects for the deuteron. 

\section{Results on the Cronin Effect}

Including these effects one can calculate the invariant cross section 
for an $AA'$ collision. Thus, introducing the nuclear modification factor
$R_{AA'}$, as
\be
\label{rdau}
R^{\pi}_{AA'} =  \frac{\dd \sigma_{\pi}^{AA'}/ \dd ^3p \,\,\, 
                (\textrm{``included nuclear effects''})}
                {\dd \sigma_{\pi}^{AA'}/\dd ^3p \,\,\, 
                (\textrm{``no nuclear effect''})} \,\, ,
\ee
nuclear effects can be investigated clearly and efficiently, on a 
linear scale. 

Following\cite{Cron75,Antr79} for the nuclear modification factor, in Figure 1, 
we present the Cronin effect as a ratio of $pW$ to $pBe$ collisions 
($R_{W/Be}$). 
This ``type'' of nuclear modification factor, is principally insensitive to
 the shadowing 
at CERN SPS energies. We compared experimental data\cite{Cron75,Antr79,e605}, with 
theoretical calculations at $\sqrt s = 19.4$, $23.8$, $27.4$ and $38.8$ AGeV 
c.m. energies. In all calculation we choose $\k2av \approx 2$ GeV$^2$, 
$\nu_{m}=4$ and $C_{sat}=0.4$ GeV$^2$ for all $A$, while 
the scale was fixed as $\kappa = 1/2$ in case of LO calculations 
({\sl dotted lines}). In NLO case the scales are chosen as $\kappa = 2/3$. 
It seems there is no difference (within $\sim 5 \%$) between curves with no 
shadowing ({\sl solid lines}) and with the updated HIJING 
$b$-independent ({\sl dashed}) or $b$-dependent ({\sl dash-dotted}) lines.  

Figure 2 presents the dependence of $R_{pA}$ ({\sl left panel}) 
and $R_{A/Be} $ ({\sl right panel}) on different targets $A \in $ 
\{ $^9 Be$, $^{32}S$, $^{44}Ti$, $^{98}In$ and $^{197}Au$ \} at CERN 
SPS energies in NLO calculations at $E_{beam} = 158$ AGeV, with $\k2av = 1.9$ 
GeV$^2$ . We present our results with EKS shadowing ({\sl dotted lines}), 
the $b$-independent, and the $b$-dependent updated HIJING shadowings 
({\sl solid lines}). The $b$-independent and $b$-dependent updated HIJING 
cases are coincide within errors. 

In our model the position of Cronin peak does not depend explicitely 
on the c.m. energy, however, depends on the value of $\k2av$. With the above 
scale choices the peak is appears at $p_T \approx 4 $ GeV. Furthermore, the 
height of the peak depends on the $A$ via $ C \cdot h_{pA}(b)$ term in eq. 
\ref{ktbroadpA}. (see Fig. 1). In the energy range
$20$ GeV $\leq \sqrt s \leq 40$ GeV shadowing effects are generally small.
Thus, the influence of the different shadowings on the Cronin effect 
are negligible, especially in minimum bias cases (see Fig. 2).  

In Figure 3 we display our NLO results for the nuclear modification 
factor in $dAu$ collision for $b$-independent ({\sl solid line}) and 
$b$-dependent ({\sl dashed line}) shadowing parameterizations from the 
updated HIJING shadowing using parameters ($\sqrt s = 200$ AGeV, 
$\k2av = 2.5$ GeV$^2$, $\kappa=4/3$).  In spite of the very different 
impact parameter dependence of $R_{dAu}$ the minimum bias results are 
very close to each other. This is because many details are averaged 
out in minimum bias data, and the final result is no longer sensitive 
to the $b$-dependence of shadowing. As we can see on Fig. 3 the 
theoretical results reproduce quite well the minimum bias experimental 
data on $\pi^0$ from PHENIX\cite{phenix03}. In minimum bias case we do not
see a strong Cronin peak. However, a reasonable Cronin peak will appear in different
centrality bins. 

Figure 4 summarizes our results on the impact-parameter ($b$) dependence 
of the Cronin effect in $dAu$ collision~\cite{dAu}. 
The nuclear modification factor, $R_{dAu}(b)$, is presented with 
updated HIJING shadowing. The {\sl left} 
column shows the $b$-independent cases, while the {\sl right} column 
displays the $b$-dependent ones. 

Applying $b$-independent shadowing for central collisions ({\sl left 
upper panel}) the curves overlap, since in these bins, the target matter is 
approximately constant from the point of view of multiscattering and 
shadowing. In more peripheral cases ({\sl left lower panel}), shadowing 
starts to dominate, $R_{dAu}(b)$ is decreasing from bin to bin. 
The $b$-independent shadowing suppresses $R_{dAu}(b)$ under 1. 
In case of $b$-dependent shadowing at central bins ({\sl right 
upper panel}) Cronin effect wins against suppression. 
At peripheral bins ({\sl right lower panel}) -- where Cronin effect 
begin to vanish and shadowing suppression working also --, the nuclear 
modification factor becomes unity.

\section{Summary }

We investigated $pp$ and $pA$ experimental data on pion 
production with our $k_T$-augmented pQCD parton model at NLO level in the
c.m. energy window $20$ AGeV $\leq \sqrt s \leq 200$ AGeV. In the lower
limit of this energy range ($\sqrt s \leq 40$ AGeV), the pion $p_T$-spectra 
is not sensitive to the applied shadowing mechanism, because shadowing is
close to be negligible ($\sim 5 \%$) in this kinematical region. 
However the multiscattering strongly influences this spectra and dominates
every deviation of $pA$ collision from superimposed nucleon-nucleon collisions  
We determined the most important 
parameters of the parton model from existing experimental data and we predicted 
the Cronin effect in $pA$ collisions at CERN SPS energy ($E_{beam}=158$ AGeV).   
Considering different targets ($^9 Be$, $^{32}S$, $^{44}Ti$, $^{98}In$ and 
$^{197}Au$) we obtained an increasing multiscattering 
effect with increasing size of the target nucleus, while the position of 
the Cronin peak did not change.   

Our analysis was repeated at RHIC energy ($\sqrt s = 200$ AGeV).
Here the shadowing effect is much larger, thus the differences between 
the competing shadowing models are emphasized. Applying $b$-dependent 
shadowing description for the $dAu$ collision the centrality dependence
of the Cronin effect generates an unusual structure in the nuclear modification
factor, which can be measured experimentally. In this way we can distinguish
between different shadowing models from the data. However minimum bias 
calculations yields similar results for different shadowing models. 
Using our model we successfully reproduced the measured nuclear 
modification factor for $\pi^0$ in the $dAu$ collision.

\section*{Acknowledgments}

This work was supported in part by Hungarian grants: T034842, T043455, T043514;
U.S. DOE grant: DE-FG02-86ER40251, and 
NSF grant: INT-0000211 Supercomputer time provided by BCPL in Norway
and the EC -- Access to Research Infrastructure action of the Improving
Human Potential Programme is gratefully acknowledged.

%%%%%%%%%%%%%%%%%%%%%%%% BIBLIOGRAPHY %%%%%%%%%%%%%%%%%%%%%%%%%%%%%%%%%%

%%%%%%%%%%%%%%%%%%%%%%%%%%%%%%%%%%%%%

\newpage
%\vspace*{-1.2cm}
\begin{center}
\vspace*{18.0cm}
\includegraphics{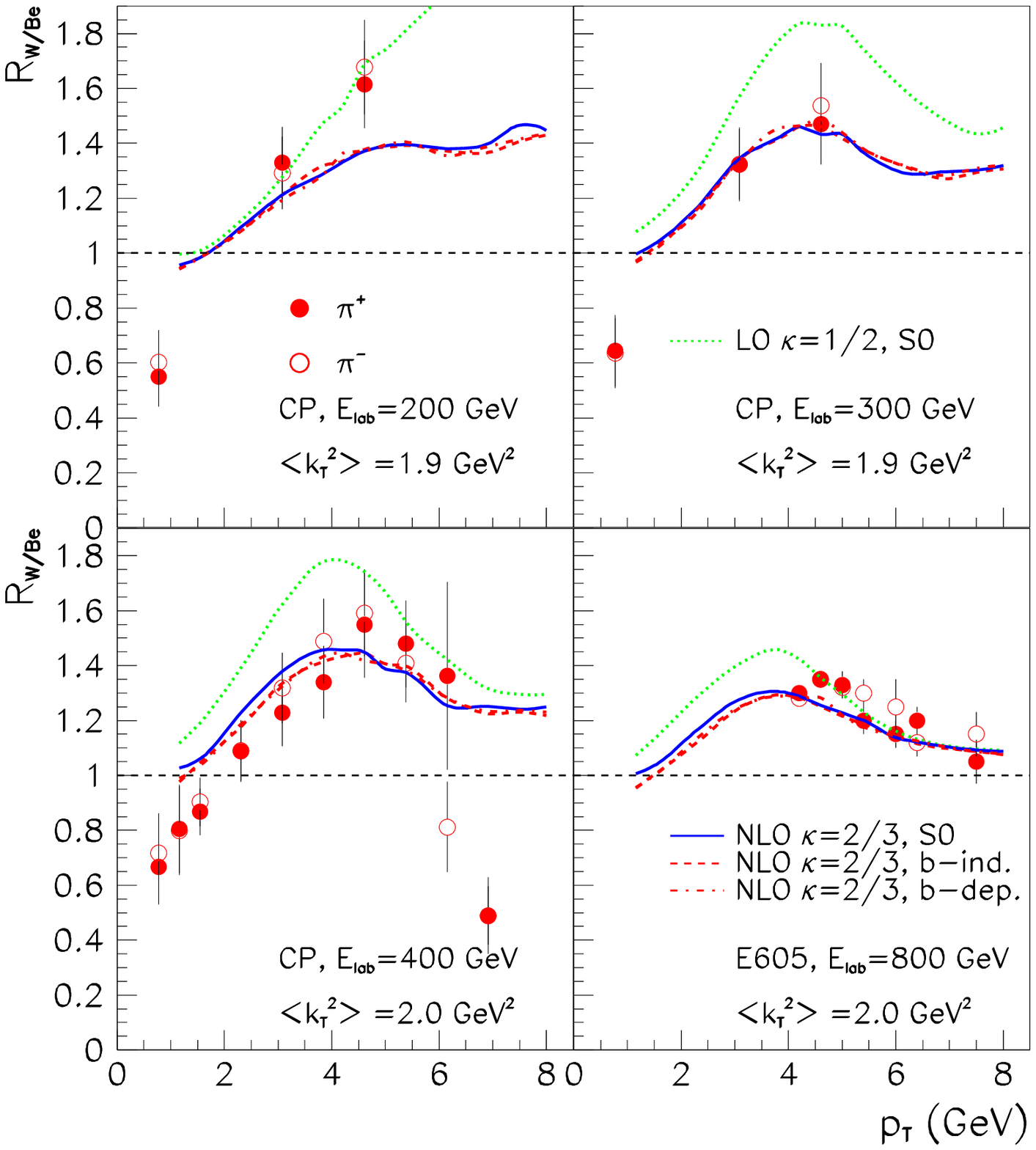}
%\vspace{-9.7cm}
\end{center}
\begin{center}
\begin{minipage}[t]{15cm}  
      { FIG. 1.} {\small Cronin effect on $R_{W/Be}$ nuclear modification
factor at CERN SPS energies, from $\sqrt s = 19.3$ AGeV up to $38.8$ AGeV
compared to experimental data\cite{Cron75,Antr79,e605}.  
Curves are in LO with {\sl dotted}, in NLO with no shadowing {\sl (S0)} with 
{\sl solid}, and
in NLO with updated HIJING shadowing\cite{Shadxnw_uj} both in 
$b$-independent {\sl dashed} and $b$-dependent {\sl dash-dotted lines} cases.
The later two are not different within errors. 
All calculations were carried out using $\nu_{m}=4$, $C_{sat}=0.4$ GeV$^2$,
and with $\langle k_T^2 \rangle$ shown. We applied
scales: $\kappa=1/2$ in LO and $\kappa=2/3$ in NLO.  
} 
\end{minipage}
\end{center}

%%%%%%%%%%%%%%%%%%%%%%%%%%%%%%%%%%%%%

\newpage
%\vspace{1.2cm}
\begin{center}
\vspace*{8.0cm}
\includegraphics{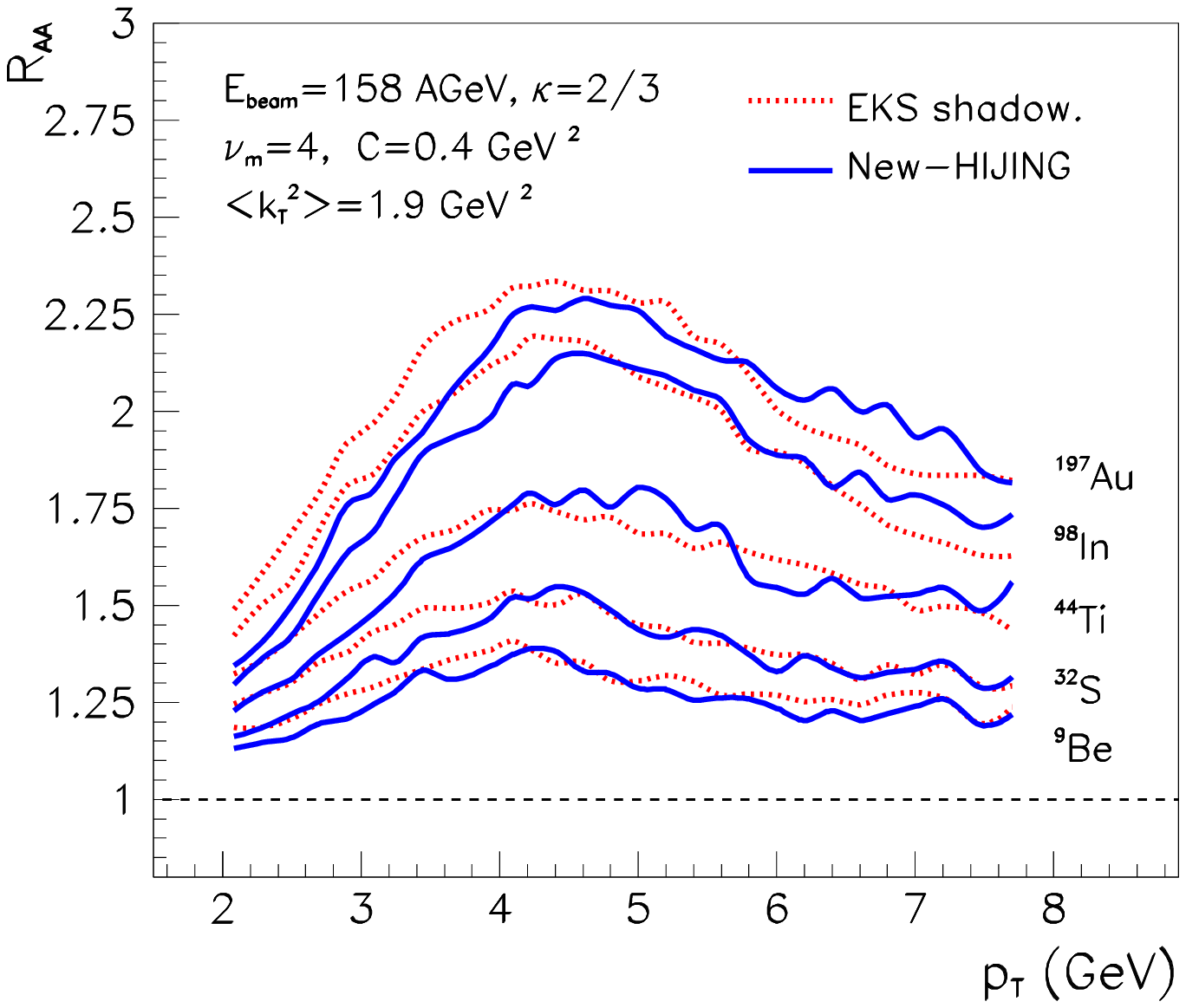}
\includegraphics{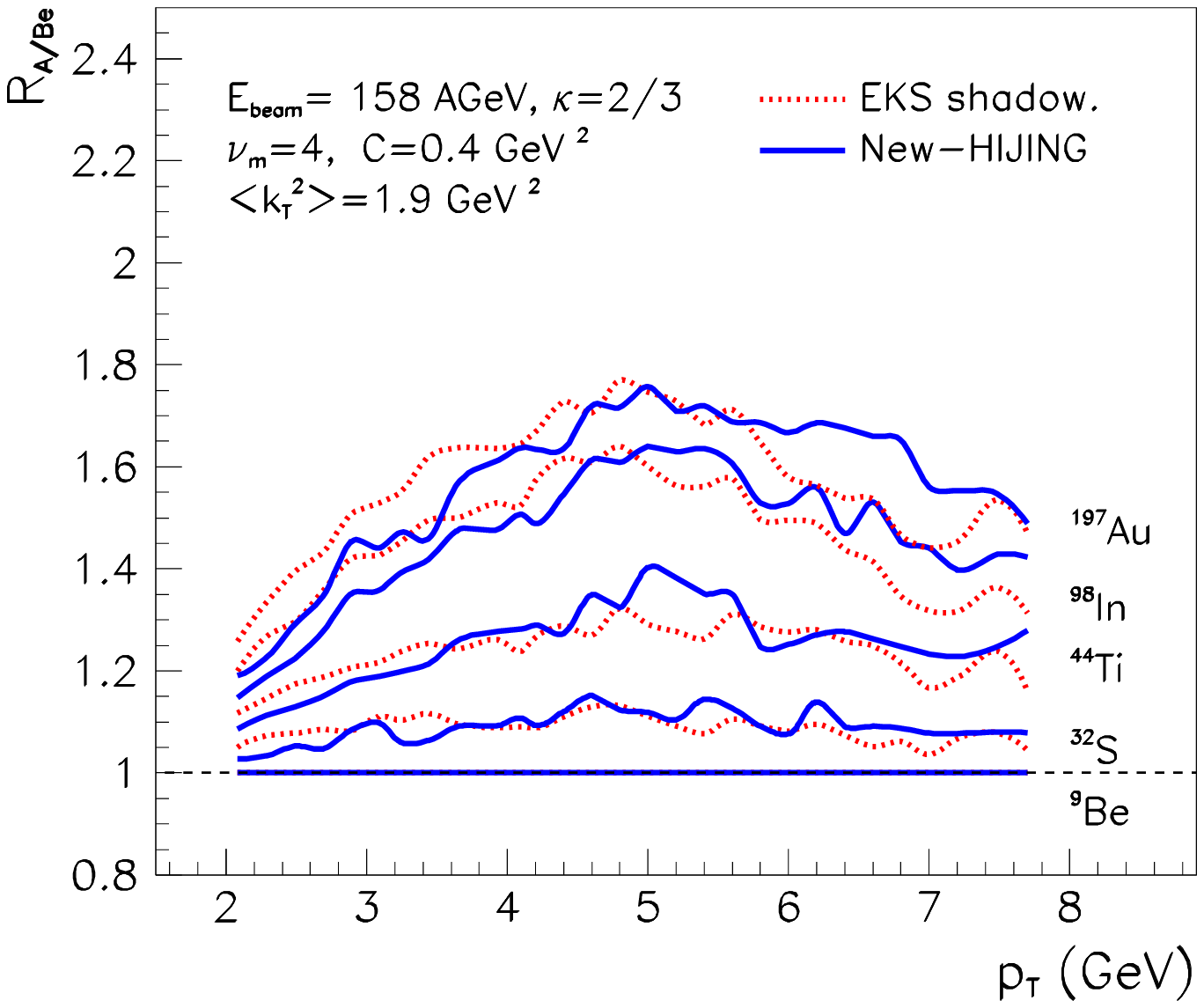}
%\vspace{-0.7cm}
\end{center}
\begin{center}
\begin{minipage}[t]{15cm}  
      { FIG. 2.} {\small 
Dependence of $R_{AA'}(p_T)$ ({\sl left panel})
and $R_{A/Be}(p_T) $ ({\sl right panel}) on different $A$ targets ($A \in $
\{$^9 Be$, $^{32}S$, $^{44}Ti$, $^{98}In$ and $^{197}Au$\}) at CERN
SPS energies. We presented our results on three types of shadowing: EKS by
Eskola et {\it al.}\cite{EKS} ({\sl dotted lines}), and
new HIJING\cite{Shadxnw_uj} shadowing ({\sl solid lines}).
The NLO calculations were carried out at $E_{beam} = 158$ AGeV,
$\k2av = 1.9$ GeV$^2$ and at $\kappa=2/3$.
} 
\end{minipage}
\end{center}

%%%%%%%%%%%%%%%%%%%%%%%%%%%%%%%%%%%%%

%\newpage
%\vspace{1.2cm}
\begin{center}
\vspace*{10.0cm}
\includegraphics{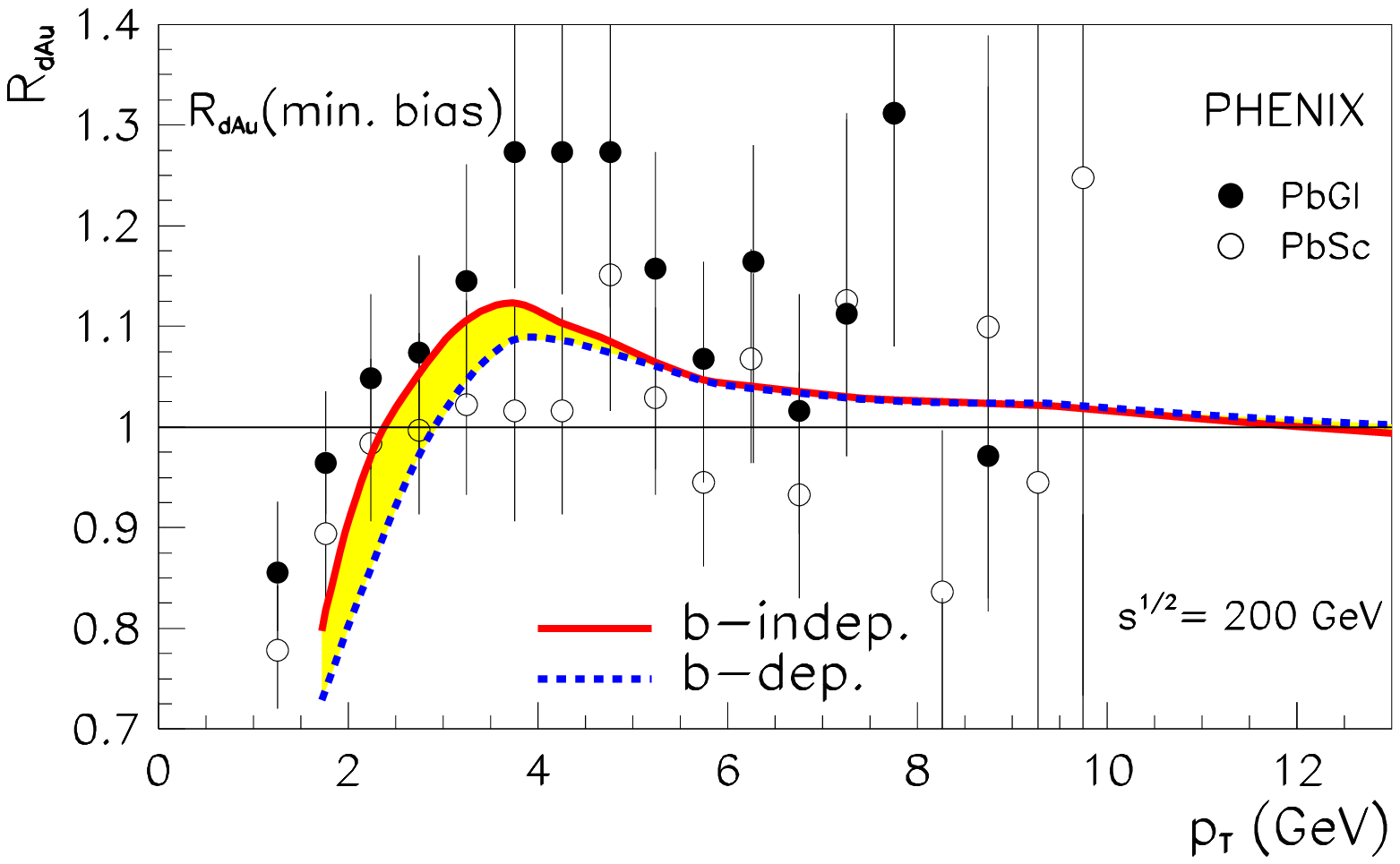}
%\vspace{-8.0cm}
\end{center}

\begin{center}
\begin{minipage}[t]{15cm}  
      { FIG. 3.} {\small The nuclear modification factor
$R_{dAu}$ for minimum bias $dAu$ collisions obtained from
our NLO calculation in case of updated HIJING $b$-independent, ({\sl solid line})
and $b$-dependent shadowing ({\sl dashed line}). Theoretical calculations are 
compared to two different experimantal sets PbGl ({\sl dots}) and 
PbSc ({\sl open circles}) at PHENIX~\cite{phenix03}.  
}
\end{minipage}
\end{center}

%%%%%%%%%%%%%%%%%%%%%%%%%%%%%%%%%%%%%

%%%%%%%%%%%%%%%%%%%%%%%%%%%%%%%%%%%%%

\newpage
%\vspace{1.2cm}
\begin{center}
\vspace*{18.0cm}
\includegraphics{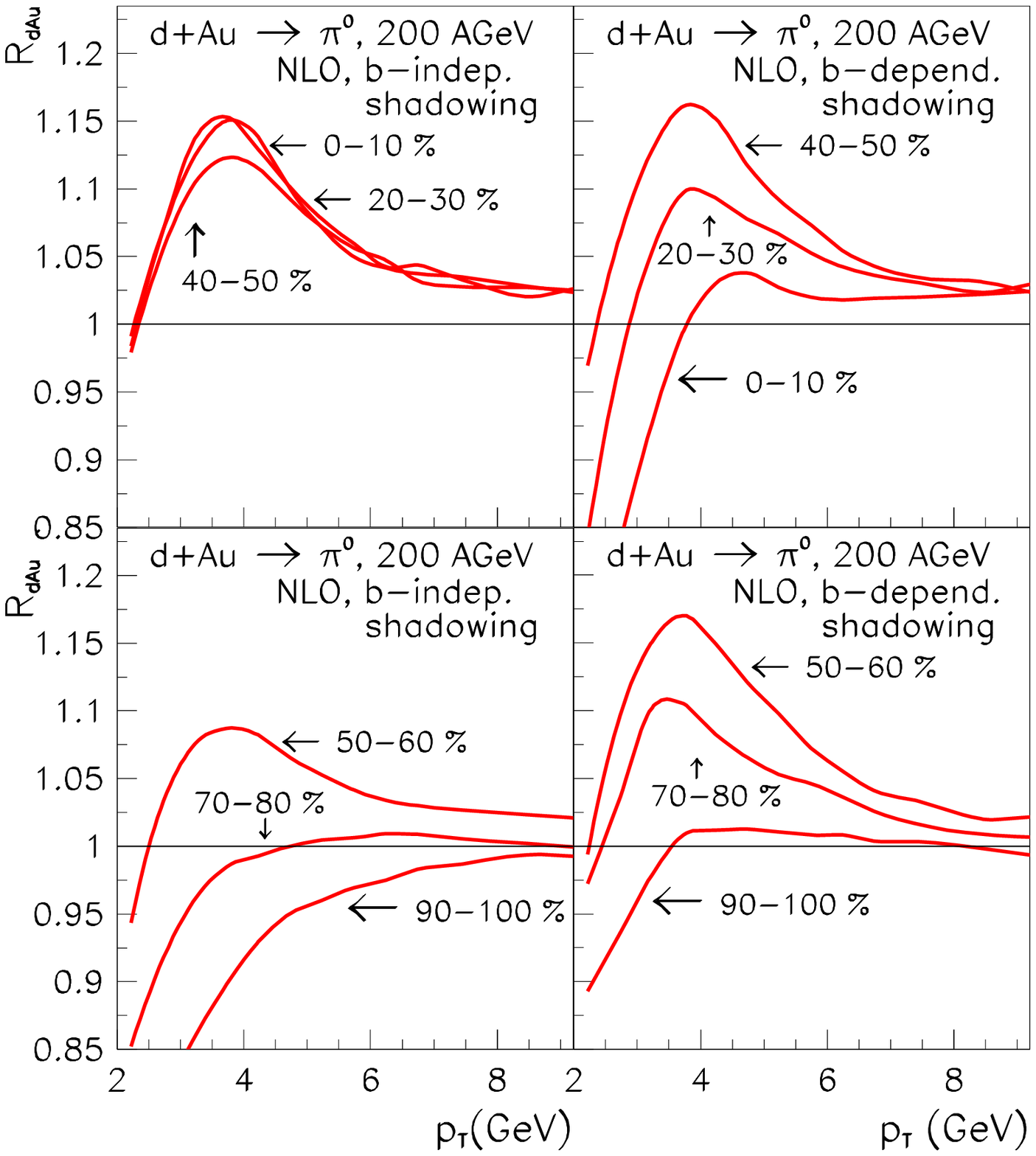}
%\vspace{-0.7cm}
\end{center}
\begin{center}
\begin{minipage}[t]{15cm}  
      { FIG. 4.} {\small The impact parameter ($b$) dependence of the
nuclear modification factor $R_{dAu}(b)$ with $b$-independent
shadowing ({\sl right column}) and with $b$-dependent shadowing ({\sl 
left column}).  The lines correspond to different centrality
bins. The $3$ ``central'' curves 
(0-50\%) are shown in the {\sl upper panels}, and the remaining $3$ 
``peripheral'' cases (50-100\%) can be seen in the {\sl lower panels}.
} 
\end{minipage}
\end{center}

\end{document}